# Physics-based Machine Learning Discovered Nano-circuitry for Nonlinear Ion Transport in Nanoporous Electrodes

*Hualin Zhan[1,2], Richard Sandberg[3], Fan Feng[3], Qinghua Liang[1], Ke Xie[1], Lianhai Zu[1], Dan Li[1,*], Jefferson Zhe Liu[3,*]*

[1] Department of Chemical Engineering, The University of Melbourne, VIC 3010, Australia

[2] School of Engineering, The Australian National University, ACT 2601, Australia

[3] Department of Mechanical Engineering, The University of Melbourne, VIC 3010, Australia

**Corresponding Author**

Dan Li: dan.li1@unimelb.edu.au

Jefferson Zhe Liu: zhe.liu@unimelb.edu.au

ABSTRACT

Confined ion transport is involved in nanoporous ionic systems. However, it is challenging to mechanistically predict its electrical characteristics for rational system design and performance evaluation using electrical circuit model due to the gap between the circuit theory and the underlying physical chemistry. Here we demonstrate that machine learning can bridge this gap and produce physics-based nano-circuitry, based on equation discovery from the modified Poisson-Nernst-Planck simulation results where an anomalous constructive diffusion-migration interplay of confined ions is unveiled. This bridging technique allows us to gain physical insights of ion dynamics in nanoporous electrodes, such as the non-ideal cyclic voltammetry.

## Introduction

Understanding and manipulation of ion transport and storage in nanoporous electrodes are essential to the development of ionic systems for applications including electrical energy storage, capacitive deionization, ionic circuits, and neuron stimulation[1-5]. While the physicochemical mechanism of ion transport and storage can be understood by molecular dynamics and continuum-level simulations, experimental interpretation and design of high-performance systems are heavily assisted by electrical models[6-12]. As the electrical response of ionic systems is generally measured at the system level, circuit models can be constructed for the whole system to interpret or predict the results if all the charge transport processes involved can be justifiably represented as appropriate electrical elements. A common example is the transmission line model for porous materials where the electrical signal applied between the parallelly aligned electrical double layer and the in-pore surface is analogous to that between two parallel transmission lines[13, 14]. Due to the straightforward implementation, electrical circuit model allows highly efficient prediction of system performance, straightforward application of practical electrical input, and direct interfacing between ionic systems and external electronic circuits for various applications[15, 16].

Numerous theoretical and experimental studies have shown that ions in nanoporous electrodes can behave differently from that in the open surface and often give rise to nonlinear dynamics[17-19]. While these discoveries open up new opportunities for design of future generations of ionic systems, the conventional circuit models used for the system-level prediction have difficulty tackling nonlinear ion transport phenomena due to the complexity in describing the diffusion part of ion transport[17, 20]. This difficulty, along with the notorious reputation of circuit models in insufficient validation of the physicochemical processes of ions, substantially hinders further exploration of circuit-based rational design and analysis of ionic systems[6, 7, 10, 17, 21]. This difficulty results from the fundamental gap between the electrical origin of circuit models and the physicochemical foundation of ionic systems, particularly when nanoporous electrodes are involved [8, 17]. In nanoporous electrodes, the electrode-electrolyte interface region extends from the exterior surface of the electrode to the interior, forming a network of interconnected nanopores. Ions need to travel through this confining nanoporous network, in which they interact intimately with each other and the electrode, further hampering the circuit representation of the physicochemical processes of ion transport. However, if the nonlinear current-voltage (I-V) relation associated with confined ion transport, including diffusion, can be mathematically expressed, new physics-based circuit elements can be constructed. This can enable a quantitative strategy to evaluate how a new nanoscale scientific discovery impacts the system performance and even to guide the design of macroscopic ionic systems through a bottom-up approach.

Recently, the emerging machine learning techniques with high interpretability are being used to discover the underlying mathematical equations for understanding unknown dynamical systems[22] and to deal with complex multiscale problems in biology and electron transport[23, 24]. In this work, we will explore the potential of using machine learning to discover mathematical equations for the complex ion transport confined in nanoporous electrodes and demonstrate how this discovery can enable a new strategy to construct physics-based nano-circuits for ionic devices. This discovery will be based on the analysis of the modified Poisson-Nernst-Planck (mPNP)[25] simulation results of confined ion transport.

**Methods**

Analysis of resistive ion transport using the conventional circuit models is based on the Drude model with the assumption that only the ion movement driven by the electrical field, so-called migration, contributes to the ion flux. The influence of the ion movement caused by a gradient of ion concentration, so-called diffusion, and the time-dependent ion concentration variations occurring at the interfaces are neglected[10]. However, theoretical studies have shown that considerable interplay between the ion migration and diffusion does exist at the electrode-electrolyte interfaces[17, 20, 21], suggesting that a fundamental theory of ion transport should be selected. Additionally, when the ion concentration approaches the maximum permitted inside nanopores, the steric repulsion of ions with finite sizes can affect the ion flux[26]. These reports prompted us to revisit the basic problems associated with nanoconfined ion transport that have often been overlooked before: how does the diffusion interplay with migration under nanoconfinement? how could their interplay affect the resultant I-V relations?

We choose the mPNP theory developed by Kilic, Bazant, and Ajdari[25] to examine above problems because migration, diffusion, and steric flux are all explicitly included and the corresponding transient potential and current profiles can be readily obtained for straightforward analysis of confined ion transport and its I-V relation[10, 25] (see Supplementary Note 1.1 for the detailed discussion of two commonly used nanoscopic methods – mPNP and molecular dynamics – and why mPNP is selected for this work). In our model, two electrodes made of cascading nanoslits are assembled to form a supercapacitor under the cyclic voltammetry (CV) charging/discharging condition (**Figures 1**a and S1). The nanoslit size, *d*, varies from 1.5 to 5 nm to ensure the validity of the mPNP theory. The standard KCl electrolyte of 1.0 M concentration is used.

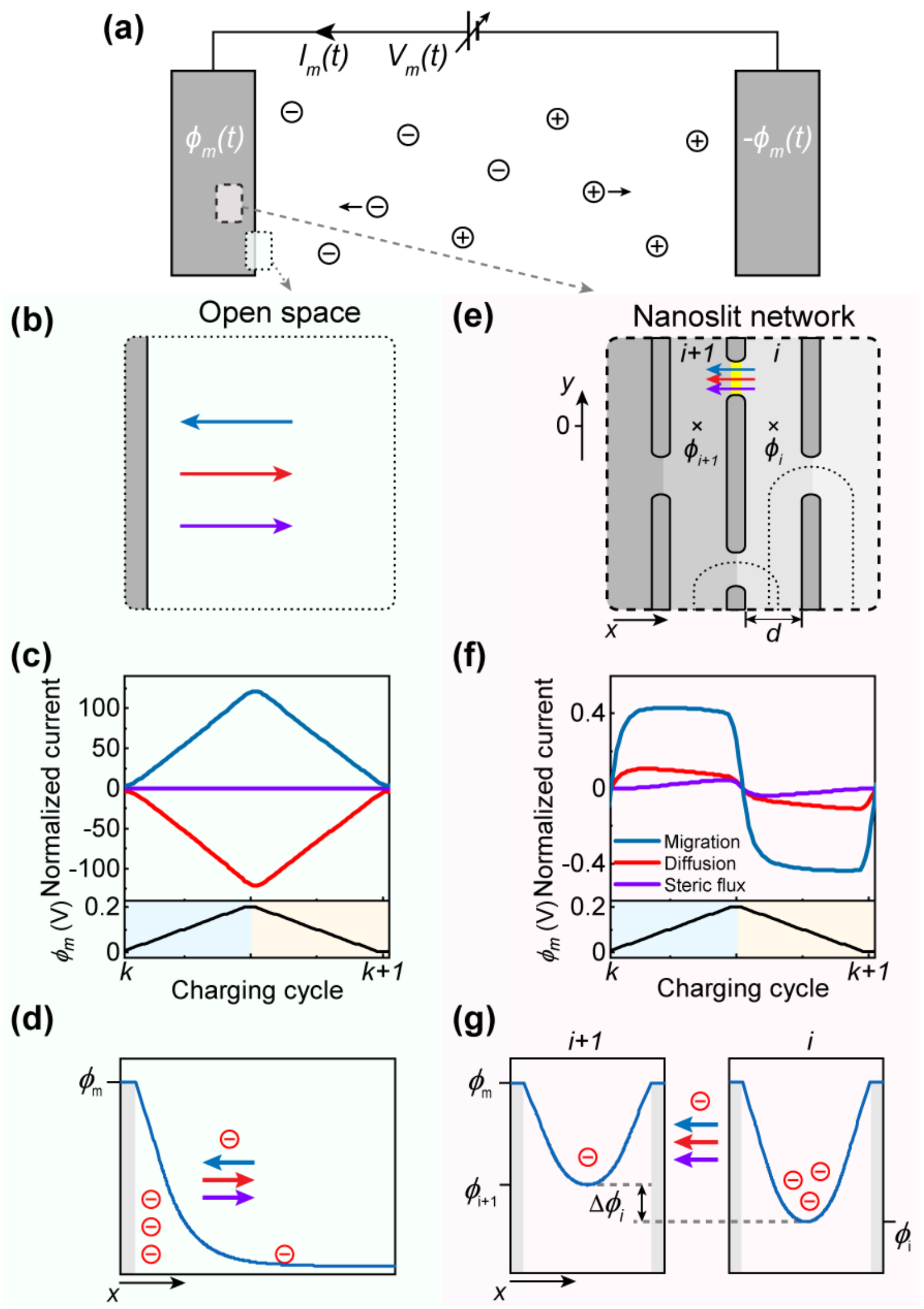

**Figure 1.** Physicochemical study of local ion transport: constructive diffusion-migration interplay in nanoslit network, as opposed to destructive interplay in open space. (a), Schematics of a supercapacitor. $I_m(t)$ is the electric current under an externally applied voltage $V_m(t)$, which splits to equal electric potential with opposite sign applied on each electrode ($V_m$=2$\phi_m$). ⊕ and ⊖ represent cations and anions. (b), The blue, red, and violet arrows indicate the direction of calculated ion migration, diffusion, and steric flux under charging in open space. (c), Calculated ionic current contributed by migration, diffusion, and steric flux during a CV cycle, which is normalized by the maximum value of $I_m(t)$. Bottom panel: variation of $\phi_m$ during charging (blue region) and discharging (yellow region). (d), Electric potential distribution in (b). (e), Nanoslits

(greyscales and indexed by $i$) are constructed between two neighbouring 2D materials (dark slats) in $x$-direction. $d$ is the slit size. The in-slit electrical potential of $i^{th}$ nanoslit ($\phi_i$) is evaluated at the marked location ($y$=0) in mPNP simulations. The dotted curves denote the EDLs of neighbouring 2D materials, which will overlap as $d$ decreases. (f), (g), Corresponding simulation results for nanoslits ($d$ =1.5 nm), as compared with (c) and (d) for open space. The current in (f) is obtained across the yellow line in (e).

**Results**

Figure 1 presents how migration, diffusion, and steric flux of net ions (i.e., the difference between cations and anions, see detailed description of mPNP in Supplementary Note 1.2) at the electrode-electrolyte interfaces in the same system behave differently when located on the exterior surface (open space, left column) and inside the nanoslit network (right column). We particularly focus on the behaviour of net ions rather than total ions because net ions are directly used in the calculation of capacitance and capacitive current for circuit models. In open space, the direction of migration (blue in Figure 1b) is observed opposite to that of diffusion (red) and steric flux (violet) during the entire CV cycle (Figure 1c and Figure S3). This is expected because when an electric field drives ions towards/away from an electrode, the increased/decreased net ion concentration on electrode surface always induce a diffusion in the opposite direction (Figure 1d)[27].

Unexpectedly, we find that the interplay between the migration and diffusion of net ions within the nanoslit network is different from that in the case of the open space. Our calculation shows the directions of the migration and diffusion of net ions from the $i^{\text{th}}$ to $(i+1)^{\text{th}}$ nanoslit within the network are the same during the entire CV cycle (Figure 1f). Such a constructive diffusion-migration interplay is also observed in the nanoslits with various $d$ values using mPNP simulations (Figure S4) and is qualitatively confirmed by our molecular dynamics simulations (Figure S5 and the Supplementary Note 1.3). We find that the direction of the steric flux is slit size-dependent; it is the same as that of migration when $d$<2 nm but becomes opposite in larger slits.

The constructive diffusion-migration interplay of net ions in the network can be attributed to the overlapping of electrical double layers (EDLs) in the nanoslits. The electrified surface attracts counter-ions to form an EDL in open space that screens the electrode potential $\phi_m$ (Figure 1d)[28]. For nanoslits with $d$ smaller than twice of the thickness of EDL, the two EDLs at the opposite surfaces overlap and form a well-shaped electrical potential profile (Figures 1g and S6a). Our analysis shows that ion transport across adjacent potential wells in neighbouring nanoslits behaves differently to that in the open space. This is because, while a large number of counter-ions in the open-space EDL corresponds to a high (un-screened) electrical potential near the electrode surface, a large number of counter-ions in a nanoslit produces a low in-slit potential. Specifically, the potential in the $i^{th}$ nanoslit is determined by the effectiveness of counter-ion screening of the electrode potential. When the number is higher in the $i^{th}$ nanoslit than that in the $(i+1)^{th}$, the more efficient screening in the $i^{th}$ nanoslit results in a lower $\phi_i$ than $\phi_{i+1}$ (Figure S6b). The resultant electric potential gradient produces a migration flux of net ion charge flux from the $i^{th}$ nanoslit to the $(i+1)^{th}$. Meanwhile, a higher net ion charge concentration in $i^{th}$ slit induces a diffusion flux in the same direction.

We find that the interplay of diffusion and migration in adjacent nanoslits produces a complex nonlinear trans-slit I-V relation over the charging/discharging process in the nanoslit network. As shown in the left panel of Figure 2a, the trans-slit voltage ($\Delta\phi_i = \phi_{i+1} - \phi_i$) does not change conformably with the trans-slit current with time ($I_i$). Such a nonlinear I-V relation is also observed when $d$ is varied (Figure S8), implying that the I-V relation cannot be described by a constant resistance. If this relationship can be expressed mathematically using ionic properties, it would enable the translation of the physicochemical knowledge of nanoconfined ion transport to electrical models.

However, the constructive diffusion-migration interplay observation does not produce mathematical equation to describe the nonlinear trans-slit I-V relation. If such an equation can be constructed, it would enable the translation of the physicochemical knowledge of nanoconfined ion transport to electrical models for efficient and quantitative prediction of dynamic charging behaviour of ionic devices. Evolutionary algorithm (EA) is particularly suitable for such an equation discovery. It, among several machine learning techniques for equation discovery including sparse regression and adjoint methods, provides a constraint-free symbolic regression method to identify equations from numerical data where no guidance is required from existing knowledge to search for the equation in a large mathematical space[29, 30].

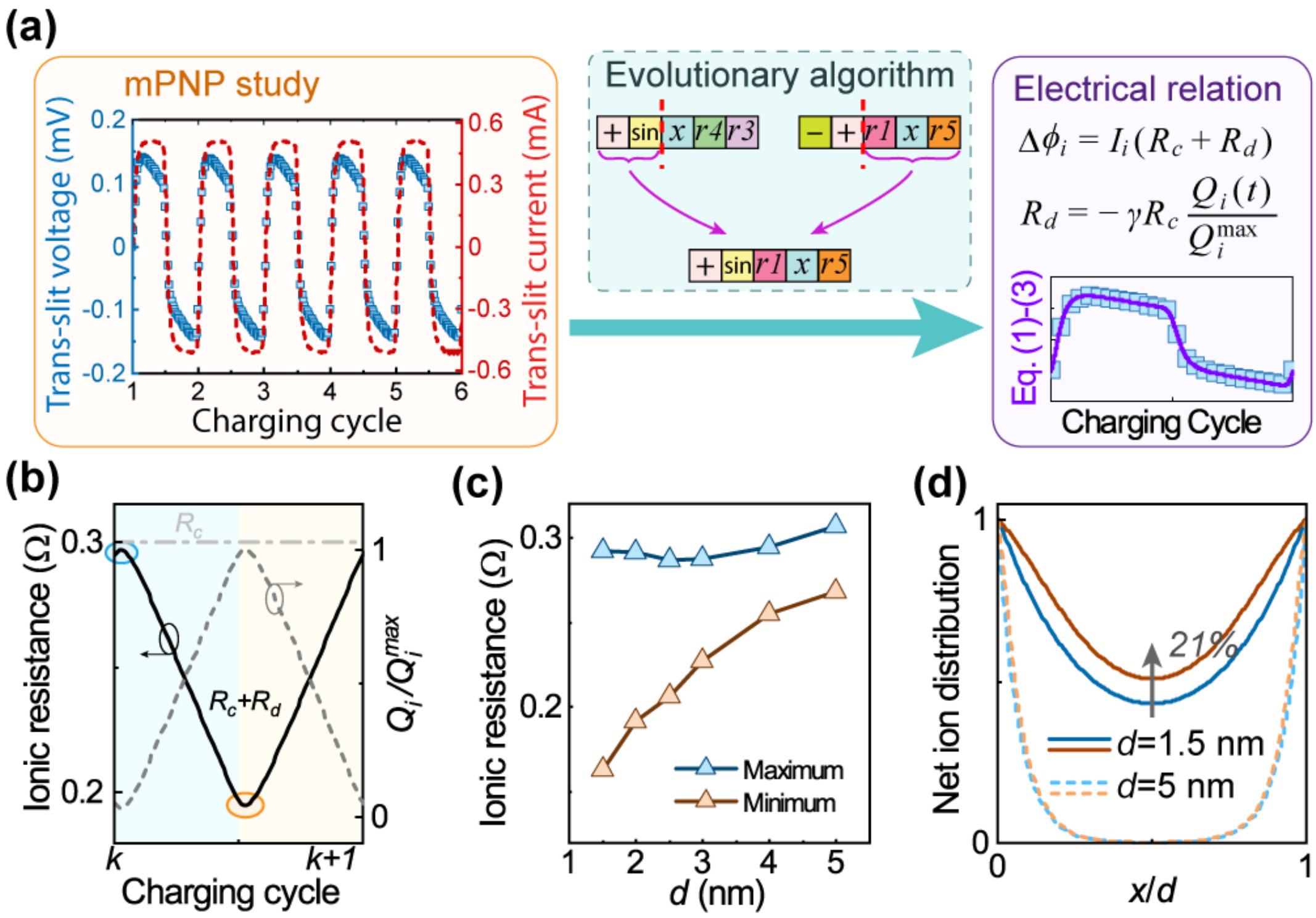

**Figure 2.** Machine-learning-discovered dynamic resistance in nanoslit network. (a), Flow chart of machine-learning-assisted equation discovery. The mPNP computational data are fed to the evolutionary algorithm (EA) analysis to yield an electrical relation including a dynamic trans-slit resistance ($R_c$+$R_d$). The cyan box shows a process in EA where a new equation for ion transport (encoded as a gene expression) is formed by taking certain sections from old equations. Inset in the purple box: EA-discovered equation (violet solid line) successfully describes the mPNP result (blue rectangles). (b), Variation of $R_c$+$R_d$ (black solid line) and net ions (grey dashed line) during a CV cycle for a nanoslit network with $d$=2 nm. $R_c$+$R_d$ reaches the maximum (circled in blue) and minimum (circled in orange) at the beginning of charging (blue region) and discharging (yellow region). (c), The maximum and minimum values of $R_c$+$R_d$ as a function of the slit size ($d$). The nearly constant maximum and the increase of minimum with the increasing $d$ are counterintuitive, as they are expected to decrease. (d), Normalized net ion distribution when $R_c$+$R_d$ reaches its maximum (blue) and minimum (orange) in nanoslits with $d$=1.5 (solid line) and 5 nm (dashed line), respectively.

The workflow of our machine-learning-discovered physical relation is summarised in Figure 2a. Here we use EA to analyse the massive data generated by mPNP simulation in order to unveil previously unnoticed mathematical/physical relations for in-pore ion dynamics. Before application of EA, we must prepare the data such that the symbolic regression could converge rapidly. As the in-pore ion dynamics is symmetric about the point where the charging process ends, *i.e.*, the centre of Fig. 1c, f, we can focus on the analysis of the charging process only. The discharging process can then be described by symmetry. However, the voltage and current are nonmonotonic functions of time, which could introduce challenges for regression convergence. This is because the algorithm would need to search in a large solution space with possibly non-elementary functions, rendering results which may not be interpretable. An observation of the simulation data suggests that the time integrations of voltage ($\int \Delta\phi_i dt$) and current ($\int I_i dt$) have a monotonic relation, where elementary functions would be sufficient for the solution space. In order to keep the integrated values within the same order of magnitude, as a common practice to facilitate regression processes[31], $\int \Delta\phi_i dt$ and $\int I_i dt$ are normalised in this work (see Supplementary Note 1.4 for detailed background of EA and algorithm implementation).

After feeding the pre-processed data to EA and allowing for thousands of evolution generations, the relation of the voltage ($\Delta\phi_i$) and current ($I_i$) between the $i^{\text{th}}$ and $(i+1)^{\text{th}}$ nanoslits in a network is identified as

$$\Delta\phi_i = I_i\, R(t) \quad (1)$$

$$R(t) = R_c + R_d(t) \quad (2)$$

$$R_d(t) = -\gamma R_c \frac{Q_i(t)}{Q_i^{max}} \quad (3)$$

Here the dynamic trans-slit resistance, $R(t)$, is composed of a constant resistance, $R_c$, and a dynamically varying resistance, $R_d(t)$. $Q_i(t)=\int I_i dt$ is the net ion accumulated in the $i^{th}$ nanoslit and $Q_i^{max}$ is its maximum. This ion accumulation is contributed by both ion diffusion and migration (Fig. S7d). $\gamma$ is a dimensionless parameter between 0 and 1, indicating how much effect does ion accumulation in nanoslit have on the ionic resistance. While $\gamma$=0 means ion accumulation in nanoslits has no effect on the ionic resistance, $\gamma$=1 describes the extreme case where a fully accumulated/occupied nanoslit has 0 ionic resistance.

## Discussion

These equations quantitatively describe the nonlinear I-V relations very well and are consistently valid when varying $d$, ion concentration, ion size, ion diffusivity, relative permittivity, and monolayer sheet size in mPNP simulations (Figures S8–S10). Exploration of how $R_c$ and $\gamma$ changes provide new perspectives and opportunities to modulate ion transport for future research; for example, we find that $\gamma$ is sensitive to electrolyte permittivity, ion concentration, and ion size, but is barely influenced by the slit length and ion diffusivity (Figures S9 and Supplementary Note 1.4). While ion diffusivity does not affect $\gamma$ or $R_d$, i.e., the ion-accumulation-induced dynamic resistance, and only changes $R_c$, small ion size tends to decrease both $R_c$ and $R_d$ even when the diffusivity is constant. These observations suggest that some electrolyte properties, such as the dielectric constant or ion size, could unexpectedly affect the trans-slit ionic resistance through the ion accumulation effect in nanoslits.

We then focus on how $R(t)$ changes with the most typical nanoconfinement condition – slit size $d$, and how this dependence could reveal new ionic conduction mechanisms in nanoslit networks. While the maximum value of $R(t)$ remains almost unchanged with $d$ (blue triangles in Figure 2c), the minimum value decreases significantly for small $d$ (yellow triangles). This appears to be counterintuitive since a smaller $d$ is expected to reduce the conductive ion numbers due to the reduced space of nanoslit and increase the value of $R(t)$, implying a different ionic conduction mechanism in nanoslits beyond this simple intuition. Our analysis shows that the ionic current across the nanoslits mainly occurs at the electrode surface (colormap in **Figure 3**a), which is analogous to the previously reported surface conduction for single nanochannels [32]. As the surface area remains constant despite the reduced nanoslit size $d$, the number of conductive ions near the surface does not change. We infer that this could account for our observation of the unchanged ionic resistance maximum which, however, does not explain the further reduced minimum ionic resistance for small $d$ alone. Further analysis shows that the net ion concentration distribution profile (orange solid curve) changes substantially after charging in the nanoslit with $d$=1.5 nm, e.g., 21% increase in the middle of the nanoslit (Figure 2d). This indicates that ions both in the centre and near the surface can contribute to ion conduction when $d$ is small. However, for $d$=5 nm the net ion distribution profile (dashed curves), which has a higher concentration near the surface, remains almost unchanged, confirming the surface conduction mechanism in large nanoslits before and after charging. This suggests distinct interactions between ions and surfaces for nanoslits with different $d$ values, i.e., the ion concentration profile in nanoslits with smaller $d$ is more sensitive to the potential or the charging status of the nanoslit network. This is likely due to the enhanced ion concentration in overlapped EDL from both nanoslit walls. Because the mPNP theory used in this work excludes other ionic processes such as re-solvation in pores smaller than 1 nm[33], equations (1)-(3) focus on the effect of overlapped EDL and do not describe these processes. Note that although nanoslits are used as an example system in this work, the overlapped EDL is not unique to nanoslits and we expect to have similar observations for the systems consisting of nanopores or nanochannels.

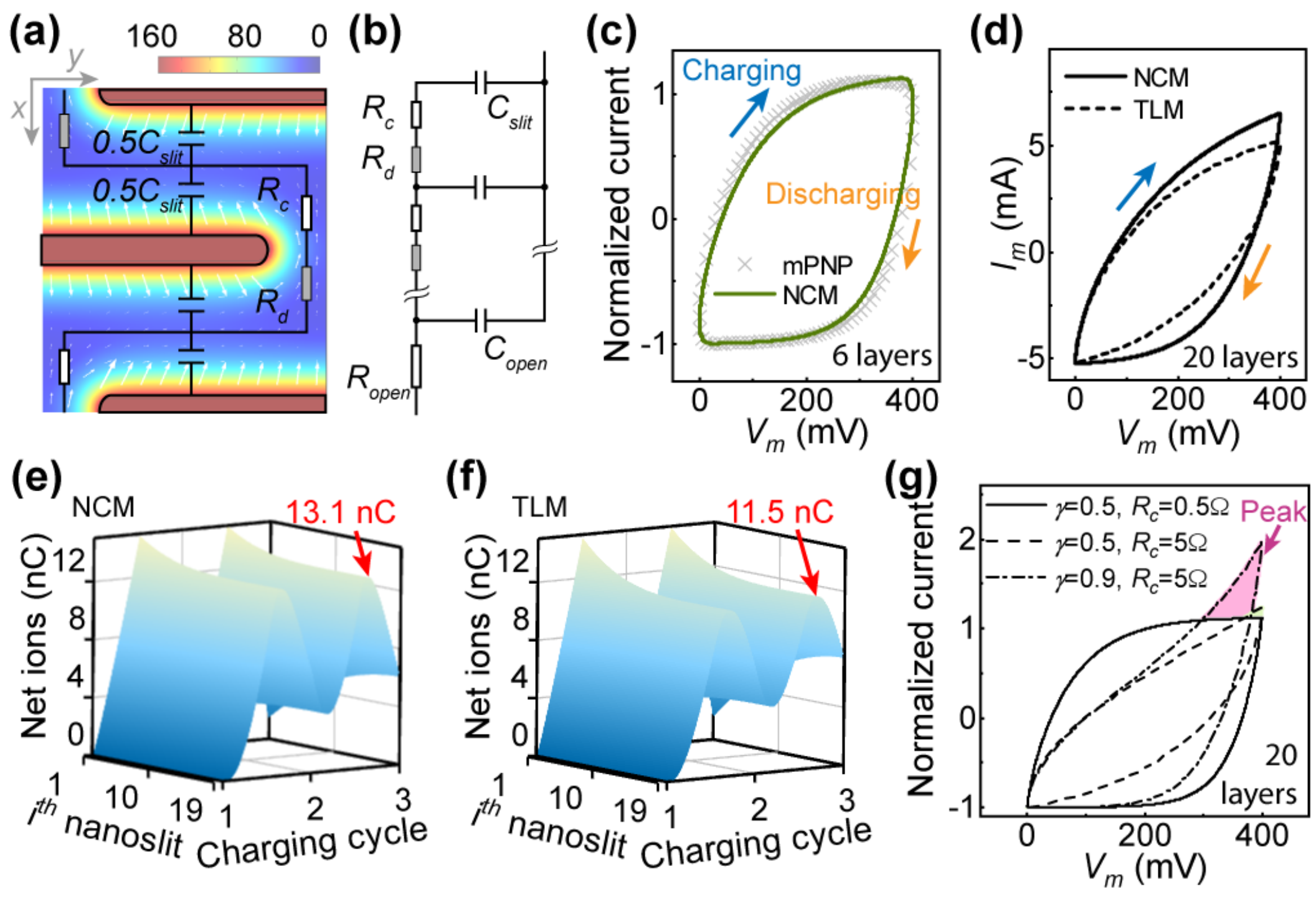

**Figure 3.** Construction of a nano-circuitry model (NCM) with machine-learning-discovered dynamic resistance. (a), Description of ion transport and storage in layered 2D material electrodes by $R_c$, $R_d$, and local capacitance of a nanoslit ($C_{slit}$), based on the spatial investigation of potential (colourmap) and current (white arrows, where the arrow size is proportional to the local current density magnitude) in nanoslits. Colour-scale unit: mV. (b), Construction of NCM to represent half-cell of a supercapacitor. (c), CV diagrams of a supercapacitor (6 layers of 2D

materials) obtained by NCM and mPNP, which agree well. (d), CV diagrams of a supercapacitor (20 layers) obtained by NCM (solid) and TLM (dashed). Note that $\gamma$=0 in NCM is equivalent to TLM. (e), (f), Net ions in the nanoslit network (20 layers, indicated by $i^{th}$ nanoslit) over two charging cycles described by NCM and TLM, respectively. Ions accumulate gradually from the open space ($1^{st}$ nanoslit) to deep inside the network ($19^{th}$ nanoslit) during charging (charging cycle 1 to 1.5 and 2 to 2.5). Here NCM captures more ion concentration inside nanoslits, as ion accumulation facilitates ion transport. In (c)-(e), $\gamma$=0.5. (g), The current peak is shown in the upper right corner as a function of $\gamma$ and $R_c$, rather than a result of redox reaction. While the increase of $R_c$ hinders ion delivery, i.e., current, at the begging of charging and makes the CV diagram less square-like, the increase of $\gamma$ facilitates ion delivery as charging proceeds and produces a peak-like CV.

Equations (1)-(3), which are expressed using ionic properties, allow us to build physics-based circuit models to predict the electrical characteristics of the entire nanoslit network. We use a supercapacitor as an example. Specifically, we allocate the nanoscopic ion process at different locations to each circuit element through investigation of the spatial distribution of electrical potential and current in the nanoslit network (Figure 3a). Ion transport across nanoslits is modelled with $R_c$+$R_d$; and the local ion storage in different nanoslit is described by a capacitor $C_{\text{slit}}$, which remains unchanged during charging/discharging when $d$>1 nm (Figure S11 and Supplementary Note 1.5). $R_{open}$ and $C_{open}$ represent the ion transport and storage in the open space of the bulk electrolyte (Figure 3b). Due to this physics-informed nature, this model can resolve nanoscale local ion transport and storage in addition to predicting the electrical response of supercapacitors, *i.e.*, nano-circuitry model (NCM). The computed CV diagrams by NCM agree well with mPNP simulations for supercapacitor electrodes consisting of 6 layers of 2D materials with $d$=1.5 nm (Figure 3c), demonstrating the validity. The number of layers can be readily increased in NCM with significantly reduced computational cost (Figure 3d), e.g., several seconds for 20 layers on a laptop computer whereas the mPNP simulations take weeks. The CV curves in Figures 3c and d capture an asymmetry between charging (blue arrow) and discharging (orange arrow), which becomes more prominent with the increased numbers of nanoslits (*i.e.*, electrode thickness) and charging rate (Figure S12). This asymmetry cannot be revealed by the transmission line model, TLM, with any parameters used (Figures 3d and S13), suggesting the critical role of $R_d$ in NCM. CV curves with trapezoid shapes are often observed in experiments [34], particularly when ions are confined in nanopores [35-38]. Desolvation of ions was previously attributed to this asymmetry [35]. Our results suggest that the asymmetry could be caused by $R(t)$ in nanoslits. The reduced trans-slit resistance facilitates ion transport and effectively delivers more ions to nanoslits for storage, as confirmed by the larger area enclosed by the solid curve than that by the dashed curve (Figures 3d and S14). This is directly observable by examining the spatiotemporal distribution of ions, where the ion concentration inside the nanoslit network predicted by NCM is higher than that by TLM (Figures 3e, f). This demonstrates an effective method approaching a desirable yet challenging aim in ionic research: simultaneous description of both the macroscopic electrical characteristics and the microscopic ionic information using cross-scale modelling methods with a low computational cost[39].

Furthermore, our NCM predicts current peaks in the CV curve, particularly when $\gamma$ and $R_c$ are large (Figure 3g and S12). Intuitive electrochemical analysis commonly attributes current peaks in CV diagram to certain Faradaic processes or redox reactions[27]. However, Faradaic processes are absent here. This again can be attributed to $R(t)$, where a substantial decrease in its magnitude leads to a much higher delivering rate of ions, *i.e.*, current, as charging proceeds. $\gamma$ could be increased by choosing small ion size, high voltage window, small $d$, low electrolyte concentration, or small dielectric constant (Figure S9). $R_c$ could be increased by lowering ion mobility, reducing electrolyte concentration, or increasing the size of the 2D material.

## Conclusions

Here we have demonstrated that machine learning can discover the underlying equation for the nonlinear ion transport in nanoporous electrodes revealed by mPNP simulations. This equation is included in a physics-based nano-circuitry model (NCM) to translate the local trans-slit nonlinearity to the electrical characteristics of the entire nanoslit network. This provides new physicochemical insights that the classic TLM fails to capture and helps reveal new mechanisms that have been overlooked by experimentalists.

Due to the limitation of the mPNP theory, the feature slit sizes are restricted to $\geq$ 1.5 nm here. Studies have shown that new ionic effects such as desolvation and ion-ion correlations can emerge when the pore size becomes smaller, and different theories need to be applied. In the future, it will be of great interest to explore how machine learning can be combined with those theories to discover nano-circuitry elements for ions under extreme confinement.

ASSOCIATED CONTENT

**Data availability.** All relevant computational results, such as the constructive diffusion-migration interplay and the generality of the machine-learning-discovered equation, are provided in the main text and the Supplementary Information. Additional data that support the findings of this study are available from the corresponding authors upon reasonable request.

**Supporting Information**. The Supporting Information is available free of charge online, which includes research methods, supplementary results, and extended discussion.